\documentclass[journal]{IEEEtran}

\usepackage{url}

\usepackage{amsmath}
\usepackage{graphicx}
\usepackage{booktabs} 
\usepackage{cite}
\usepackage{balance}

\usepackage{orcidlink}
\hypersetup{hidelinks}

\begin{document}

\title{Blind Video Denoising for Vocal Tract rtMRI using a Self-Supervised Ensemble}

\author{Julian~Andres~Hernandez-Potes\,\orcidlink{0000-0001-5994-7467},
Tomás~Arias-Vergara\,\orcidlink{0000-0001-9405-4154},~\IEEEmembership{Member,~IEEE,}
Daiqi~Liu\,\orcidlink{0009-0001-0905-3833},\\
Andreas~Maier\,\orcidlink{0000-0002-9550-5284},~\IEEEmembership{Member,~IEEE,}
and~Paula~Andrea~Perez-Toro\,\orcidlink{0000-0002-2727-2116},~\IEEEmembership{Member,~IEEE}%



\thanks{J. A. Hernández-Potes is with the German Cancer Research Center (DKFZ), 69120 Heidelberg, Germany (e-mail: julian.hernandez@dkfz-heidelberg.de).}

\thanks{J. A. Hernandez-Potes, T. Arias-Vergara, D. Liu, A. Maier and P. A. Perez-Toro are with the Friedrich-Alexander-Universität Erlangen-Nürnberg, 91058 Erlangen, Germany. (e-mail: tomas.arias@fau.de; daiqi.deutschfau.liu@fau.de; andreas.maier@fau.de; paula.andrea.perez@fau.de).}

\thanks{P. A. Perez-Toro is affiliated with the Technische Universität München, 80333 Munich, Germany.}}

\maketitle

\begin{abstract} 
Real-time magnetic resonance imaging of the vocal tract makes speech
articulation directly observable, but the trade-off between temporal resolution and signal quality introduces substantial noise and obscures fine articulatory detail. Since clean references cannot be acquired for a moving vocal tract, supervised denoising approaches are not applicable in this setting. We propose a self-supervised denoising framework that combines spatiotemporal super-resolution learning with pseudo-supervision derived from a variance-stabilized non-local means filter. A dual-model fusion strategy balances noise suppression and preservation of high-frequency articulatory structure. Experiments on the USC-75 multi-speaker dataset under natural noise conditions show that the ensemble raises the signal-to-noise ratio by 11.9 dB on unseen subjects while attaining more high-frequency articulatory content in the frames. Among tuned implementations of anisotropic diffusion, block-matching filtering, non-local means and pre-smoothed non-local means, every configuration that achieves comparable noise suppression does so at the cost of high-frequency content or of structural fidelity. Under severe added Gaussian, Rayleigh and Rician noise, peak signal-to-noise ratio improves by 7.6 to 11.8 dB and structural similarity roughly triples. An ablation shows that the variance-stabilized pseudo-supervision is an essential element for denoising, and that the effect is attributable to the hard thresholding contained in the stabilization transform.
\end{abstract}


\section{Introduction}

\IEEEPARstart{M}{agnetic} resonance imaging (MRI) data are inherently susceptible to noise, particularly in real-time, fast, or experimental sequences, which degrades image quality~\cite{rician}. Whereas denoising methods have been widely developed for brain, fetal and cardiac imaging, analogous efforts for the vocal tract remain relatively underexplored~\cite{survey,Ramanarayanan2018}. Such limitation is particularly relevant for real-time MRI (rtMRI) of the vocal tract, where the trade-off between temporal resolution and signal-to-noise ratio (SNR) produces noisy magnitude frames that obscure fine articulatory structures, such as the tongue and velum, which are essential for accurate speech analysis.

With the rise of deep learning, MRI denoising has shifted towards
learning-based schemes. As reported in the reviews by Chen et al.~\cite{survey} and more recently by Nazir et al.~\cite{nazir2024}, convolutional encoder-decoder architectures such as the U-Net have become the standard and account for the majority of the published work along with transformer-based models, offering superior capability in capturing global dependencies and preserving fine structural details amidst complex noise patterns, such as Rician noise. Nevertheless, a recurring limitation identified in the current literature is the heavy reliance on supervised training with large paired datasets. While effective for static anatomy ~\cite{wgan}, this dependence on clean targets presents a fundamental bottleneck for dynamic speech imaging, where physiological motion renders the acquisition of noise-free reference data physically and practically unfeasible.

Recent research instead exploits the internal correlations of high-dimensional acquisitions to build supervisory signals without ground truth. N2SR-M by Tian et al.~\cite{n2srm} trains on spatially sub-sampled pairs from noisy data to restore fine structures via Super-Resolution (SR) constraints. Although it was formulated for multi-contrast MRI, the principle transfers to the temporal dimension of rtMRI, and it has been validated in quantitative MRI sequences~\cite{kang2024} and in high-speed fluorescence imaging~\cite{wang}. Complementarily, we believe that conventional analytical approaches remain indispensable for their explicit handling of noise statistics and spatial signal characteristics~\cite{nlm, li}. Works such as that of Yang et al.~\cite{brainriciannlm} have addressed Rician noise reduction via a Pre-Smoothed Non-Local-Means (PSNLM) filter with variance stabilization, demonstrating that explicitly modeling noise characteristics is crucial for preserving anatomical detail in low-SNR regimes.

Motivated by these complementary strengths, we propose a novel self-supervised ensemble for vocal tract rtMRI denoising that combines the spatiotemporal SR strategy of N2SR-M with the noise stabilization of PSNLM. Unlike single-frame blind-spot networks~\cite{n2void, Xu_2020, flourescence} that rely solely on spatial context, our method exploits the spatiotemporal redundancy of speech sequences to disentangle stochastic noise from deterministic articulatory motion, achieving robust denoising without the need for a clean ground truth.

\begin{figure*}
    \centerline{\includegraphics[width=0.90\textwidth]{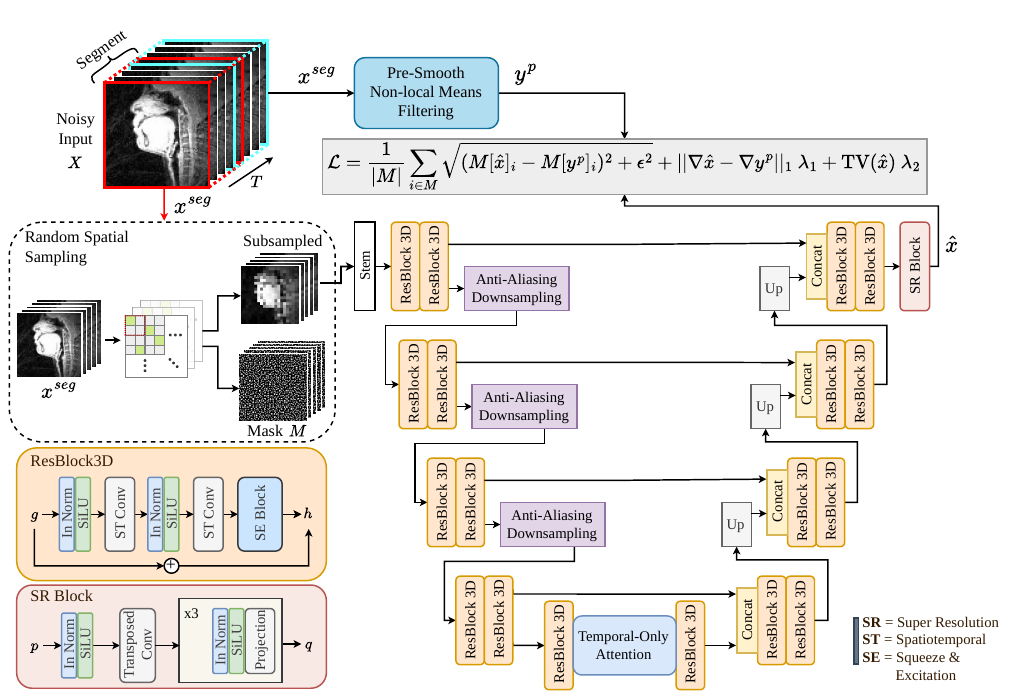}}
    \caption{Training framework for our denoising network. The noisy rtMRI video is cut into overlapping clips; the spatial sub-sampling block retains one pixel from every $2\times2$ cell, which forms the network input, and marks the remaining three quarters in the mask $M$. The network, detailed in Sec.~\ref{subsec:c}, is optimized against a PSNLM-filtered version $y^{p}$ of the same clip with a composite loss that preserves edges and limits over-blurring. Only the fidelity term is restricted to $M$. Eq.~\eqref{eq:loss} defines all three terms of the loss function.}
    \label{fig:model}
\end{figure*}

\section{Dataset} \label{sec:2}

We use the USC-75 dataset~\cite{usc75}, a multi-speaker public collection of raw and reconstructed 2-D sagittal rtMRI videos capturing speech production at the vocal tract in 75 healthy speakers. The reconstructed magnitude videos have a spatial resolution of $84\times84$ pixels and a temporal resolution of 83~frames/s. Consecutive frames are strongly correlated~\cite{Lingala2015,book}, so we decimate the temporal axis by a factor of~4, which reduces the computational cost while preserving articulatory events at the time scale of interest. Rician noise and reconstruction artifacts are present in every available magnitude frame, and no noise-free counterpart exists.

The proposed networks were trained on a pool of 54 subjects, and all results reported below are measured on subjects outside that pool. Four clips from four unseen subjects reading the \textit{Grandfather} passage are reserved for selecting the hyper-parameters of the analytical filters, so that no baseline is tuned on the data used for its evaluation. The remaining 17 unseen subjects, all reading the \textit{North Wind} passage, form the test set. 


\section{Methods}

We formulate rtMRI denoising as a self-supervised problem where no clean ground truth exists. Let $X \in \mathrm{I\!R}^{T\times H \times W}$ denote a magnitude video corrupted by signal-dependent noise, segmented into the overlapping clips $x^{seg}$ of Fig.~\ref{fig:model} before being fed to the network. 

\subsection{Spatiotemporal Self-Supervision} \label{subsec:ss}
As illustrated in Fig.~\ref{fig:model}, we adopt a similar sub-sampling strategy as N2SR-M~\cite{n2srm} to enable training without noise-free targets. The operator $\mathcal{D}_s: \mathrm{I\!R}^{T\times H\times W} \rightarrow \mathrm{I\!R}^{T\times \frac{H}{2}\times \frac{W}{2}}$ partitions every frame into non-overlapping $2\times2$ cells and keeps one pixel per cell, drawn uniformly at random and independently for each cell and each frame. The retained pixels form the network input, while the remaining three quarters, collected in the mask $M$, are the only pixels on which the fidelity term of the loss is evaluated. Since the withheld pixel is not part of the input, its noise realization cannot be predicted from the observations, and the network is constrained to interpolate the underlying anatomy rather than to reproduce the noise of the target. The network $f_\theta$ is thus trained to recover the full-resolution structure under the overall objective:
\begin{equation}
    \hat{\theta} = \arg\min_\theta\; 
    \mathrm{I\!E}_{X}\!\left[\mathcal{L}\!\left(f_\theta\bigl(\mathcal{D}_s(X)\bigr),\; 
    y^{p}\right)\right],
    \label{eq:obj}
\end{equation}
where $y^p$ is the pseudo-target defined in the following subsection. Because the sampled positions differ from frame to frame, reconstructing the withheld pixel requires not only aggregating information from spatial neighbors but also from adjacent frames.


\subsection{Variance-Stabilized Pseudo-Supervision} \label{sec:3b}

In a magnitude image the noise follows a Rician distribution. For an underlying signal amplitude $A$ and Gaussian noise of standard deviation $\sigma$ in each of the two orthogonal components of the complex-valued MRI signal:

\begin{equation}
    M_{\mathrm{obs}}=\sqrt{(A+n_1)^2+n_2^2},\quad n_1,n_2\sim\mathcal{N}(0,\sigma^2),
    \label{eq:rice}
\end{equation}
so that $\mathrm{I\!E}[M_{\mathrm{obs}}^2]=A^2+2\sigma^2$ and both the mean and the variance depend on the underlying SNR $r=A/\sigma$. In air and inside the airway, where $A\approx0$, the magnitude follows a Rayleigh distribution with mean $\sigma\sqrt{\pi/2}$. The noise in these regions does not have zero mean, so a raw noisy frame is a biased regression target in the regions that contain the articulatory structures of interest. Each frame is therefore filtered in a domain in which the noise is approximately additive and homoscedastic, and the result is used as the regression target:

\begin{equation}
    y^p_t = F^{-1}\!\left(\mathrm{PSNLM}\!\left(F(X_t)\right)\right),\quad
    t=1,\dots,T,
    \label{eq:pseudo}
\end{equation}
with $F$ a variance-stabilizing transform (VST), $F^{-1}$ its inverse and PSNLM a pre-smoothed non-local means filter. 

The choice of $F$ is not established for magnitude MRI, so two transforms are considered and compared in the ablation of Sec.~\ref{subsec:abl}. The first, $F_{\mathrm{V}}$, is the transform of Foi~\cite{foi2011} obtained by quadrature over the Rice density so that the transformed data have the same standard deviation at every SNR, and inverted with the exact unbiased inverse. In our implementation the output standard deviation deviates by at most 0.8\,\% from a constant over $0\le r\le12$, compared with 34\,\% for the untransformed magnitude. Its noise level is estimated per frame from the magnitudes $M_{\mathrm{bg}}$ inside a background rectangle, where $A\approx0$ and the Rayleigh median gives $\hat{\sigma}=\mathrm{median}(M_{\mathrm{bg}})/\sqrt{2\ln 2}$. The second, $F_{\mathrm{T}}$, is a closed-form algebraic approximation and is the transform used to generate the pseudo-targets of the models reported here. It is applied pixel by pixel and combines a Gudbjartsson--Patz~\cite{rician} bias correction with an Anscombe-type root:

\begin{equation}
    F_{\mathrm{T}}(M_{\mathrm{obs}}) = 2\sqrt{\max(M_{\mathrm{obs}}^{2}
    - \hat{\sigma}_{\mathrm{O}}^{2},\, 0) + \tfrac{3}{8}},
    \label{eq:thresh}
\end{equation}
here $\hat{\sigma}_{\mathrm{O}}$ is the standard deviation of the pixels below an Otsu threshold, and the transform is inverted by $F_{\mathrm{T}}^{-1}(u)=\sqrt{\max((u/2)^{2}-\tfrac{3}{8} +\hat{\sigma}_{\mathrm{O}}^{2},\,0)}$.

\subsection{Network Architecture} \label{subsec:c}

We design a spatiotemporal U-Net with a pre-activation scheme~\cite{he}, consisting of four main components:
\begin{enumerate} 
\item \textsl{Spatiotemporal Feature Extraction:} the backbone consists of 3-D residual blocks enhanced with squeeze-and-excitation (SE) modules~\cite{senet} to model channel-wise dependencies.
\item \textsl{Anti-Aliasing Downsampling:} strided convolutions are replaced with non-learnable Gaussian blur filters to mitigate aliasing artifacts.
\item \textsl{SR Upsampling:} as in N2SR-M, reconstruction is performed by a SR block of 3-D transposed convolutions followed by projection-based refinement layers.
\item \textsl{Temporal-Only Attention:} a bottleneck attention block that focuses strictly on temporal dependencies to enforce frame-to-frame consistency.
\end{enumerate}

The encoder has four stages with channel multipliers $(1,2,4,8)$ and two residual blocks per stage, and the SR block outputs twice the input grid in each spatial dimension.

\subsection{Composite Loss and Dual-Model Fusion}
The network is optimized with a composite loss enforcing data fidelity, edge preservation, and smoothness. With $\hat{x}$ the network output and $M$ the mask
of pixels withheld by $\mathcal{D}_s$:

\begin{equation}
    \mathcal{L} = \mathcal{L}_{\mathrm{char}}(M[\hat{x}], M[y^{p}]) 
    + \lambda_1\,\mathcal{L}_{G}(\hat{x}, y^{p})  
    + \lambda_2\,\mathcal{L}_{\mathrm{TV}}(\hat{x}),
    \label{eq:loss}
\end{equation}
where $\mathcal{L}_{\mathrm{char}} = \frac{1}{|M|}\sum_{i \in M}\sqrt{(M[\hat{x}]_i - M[y^p]_i)^2 + \epsilon^2}$ denotes the Charbonnier penalty over masked pixels, $\mathcal{L}_{G} = \|\nabla\hat{x} - \nabla y^p\|_{1}$ compares Sobel
gradient magnitudes to promote sharp boundaries, and $\mathcal{L}_{\mathrm{TV}}$ is the total variation (TV) loss of the output to control smoothness.

We observed that a single value of $\lambda_1$ imposes a compromise between texture preservation and noise suppression, so we train two networks with different values and fuse their outputs. $f_{\mathrm{HF}}$ uses 16 base channels, 3-frame clips and $\lambda_1=1.2$, which favors high-frequency detail and edges; $f_{\mathrm{LF}}$ uses 48 base channels, 40-frame clips and $\lambda_1=0.7$, which favors global smoothness and the preservation of the original anatomical structure. The denoised video is their convex combination:

\begin{equation}
    X_{d} = w\,f_{\mathrm{HF}}(x^{seg}) + (1-w)\,f_{\mathrm{LF}}(x^{seg}),
    \label{eq:ens}
\end{equation}
which is both locally sharp and globally consistent. The weight $w$ is not fitted; it is set manually to trade off noise suppression against detail retention. We use $w=0.8$ on the original data, where the noise level is moderate, and $w=0.1$ under severe synthetic corruption.

\subsection{Training and Inference}

Both networks are trained with AdamW, a learning rate of $2\cdot10^{-4}$ for $f_{\mathrm{HF}}$ and $5\cdot10^{-4}$ for $f_{\mathrm{LF}}$, and a plateau schedule on the validation loss. At inference the sub-sampling is removed and the full-resolution video is passed through both networks; overlapping clips are merged by averaging and the super-resolved output is resampled back to the acquisition grid before the fusion of Eq.~\eqref{eq:ens}. No filter is applied at inference; all values reported for the proposed models are raw network outputs.

\begin{figure*}[!t]
    \centerline{\includegraphics[width=0.97\textwidth]{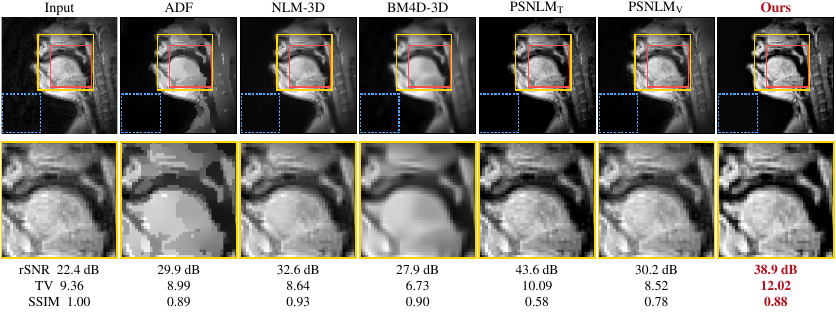}}
    \caption{One frame of the unseen \texttt{sub075} clip. The top row is the full frame, with the signal region that forms the numerator of Eq.~\eqref{eq:rsnr} (red) and the background region-of-interest that forms its denominator (blue). The bottom row displays a magnified view of the full frame (yellow); all printed metrics are 50-frame averages over the clip, not only the displayed frame.  ADF quantizes the soft tissue into piecewise-constant regions, BM4D-3D removes the noise together with the textures, and $\mathrm{PSNLM}_{\mathrm{T}}$ preserves the anatomy but flattens the background, while $\mathrm{PSNLM}_{\mathrm{V}}$ changes the signal much less. The proposed ensemble is the only denoised output whose TV exceeds that of the input while improving SNR and preserving structural fidelity.}
    \label{fig:visual}
\end{figure*}

\section{Results}

We assess performance in two scenarios: the original USC-75 rtMRI data, which contain only the noise present at acquisition, and a controlled setting in which synthetic noise is added to the same videos. In both, the ensemble is benchmarked against representative analytical denoisers, after which the pseudo-supervision that drives the training objective is ablated.

\subsection{Evaluation protocol}
On the original data no clean reference exists, so we use three reference-free measures. The first is a region-of-interest SNR:

\begin{equation} 
    \mathrm{rSNR} = 10\log_{10} \frac{\mu_{\mathcal{S}}^{2}} {\left(\hat{\sigma}_{\mathcal{N}}/\sqrt{2-\pi/2}\right)^{2}}, 
    \label{eq:rsnr} 
\end{equation} 
where $\mu_{\mathcal{S}}$ is the mean intensity over a fixed $30\times30$ vocal-tract window and $\hat{\sigma}_{\mathcal{N}}$ the standard deviation over a fixed $28\times28$ background window in the lower-left corner of the frame, both drawn in Fig.~\ref{fig:visual}. The second measure is TV, which is included because rSNR alone favors over-smoothed outputs. The third is Structural Similarity Index (SSIM) computed against the noisy input to quantify agreement with the input anatomy. For every method, the input included, 50 consecutive frames are scored per clip.

We benchmark against Perona--Malik anisotropic diffusion (ADF)~\cite{perona1990,gerig1992}, the volumetric transform-domain filter BM4D~\cite{maggioni2013}, non-local means (NLM)~\cite{nlm}, and the PSNLM filter~\cite{brainriciannlm} that supplies our pseudo-targets, the latter in both stabilizations of Sec.~\ref{sec:3b}: $\mathrm{PSNLM}_{\mathrm{T}}$ uses the thresholded $F_{\mathrm{T}}$ and $\mathrm{PSNLM}_{\mathrm{V}}$ the Rician $F_{\mathrm{V}}$. BM4D, NLM and PSNLM are additionally run in a spatiotemporal form, marked 3-D, which filters the whole $T\times H\times W$ volume. Their hyper-parameter grids were scored inside this same protocol on the tuning clips of Sec.~\ref{sec:2}. For each filter we retain the configuration that maximizes rSNR subject to $\mathrm{SSIM}\ge0.90$. Where no configuration satisfies this constraint, as is the case for $\mathrm{PSNLM}_{\mathrm{T}}$, the unconstrained rSNR optimum is reported and marked as such.

\begin{table}[!t]
\centering
\caption{Reference-free quality averaged over the 17 test subjects. [pseudo-target] is the setting that generates $y^p$, [max $r$SNR] the unconstrained optimum of the search and the unmarked rows its $\mathrm{SSIM}\ge0.90$ optimum. $f_{\mathrm{HF}}$ and $f_{\mathrm{LF}}$: high- and low-frequency components of the network respectively.}
\label{tab:reffree}
\footnotesize
\setlength{\tabcolsep}{3pt}
\begin{tabular}{lccc}
\toprule
Method & rSNR $\uparrow$ & TV & SSIM $\uparrow$ \\
\midrule
Original (noisy) & 21.5$\pm$2.4 & 9.32$\pm$0.99 & 1.00$\pm$0.00 \\
\midrule
ADF & 28.2$\pm$4.6 & 8.98$\pm$0.94 & 0.89$\pm$0.02 \\
BM4D (3-D) & 26.7$\pm$4.1 & 7.05$\pm$0.64 & 0.92$\pm$0.02 \\
NLM (3-D) & 31.6$\pm$5.9 & 8.31$\pm$0.71 & \textbf{0.95$\pm$0.04} \\
PSNLM$_{\mathrm{T}}$ [pseudo-target] & 35.7$\pm$9.7 & 9.96$\pm$1.12 & 0.57$\pm$0.04 \\
PSNLM$_{\mathrm{T}}$ [max rSNR] & 40.0$\pm$9.9 & 5.81$\pm$0.70 & 0.64$\pm$0.04 \\
PSNLM$_{\mathrm{V}}$ [pseudo-target] & 27.9$\pm$5.1 & 8.51$\pm$0.81 & 0.80$\pm$0.09 \\
PSNLM$_{\mathrm{V}}$ [tuned] & \textbf{20.5$\pm$2.4} & 9.46$\pm$1.04 & 0.92$\pm$0.03 \\
\midrule
$f_{\mathrm{HF}}$ (ours) & 32.0$\pm$7.6 & 12.73$\pm$1.33 & \textbf{0.87$\pm$0.04} \\
$f_{\mathrm{LF}}$ (ours) & 39.9$\pm$10.1 & 9.31$\pm$0.89 & 0.56$\pm$0.03 \\
Ensemble (ours) & \textbf{33.4$\pm$8.0} & \textbf{12.05$\pm$1.24} & 0.85$\pm$0.04 \\
\bottomrule
\end{tabular}
\end{table}

\subsection{Reference-Free Evaluation} \label{subsec:reffree}

\begin{figure}[!t]
    \centerline{\includegraphics[width=\columnwidth]{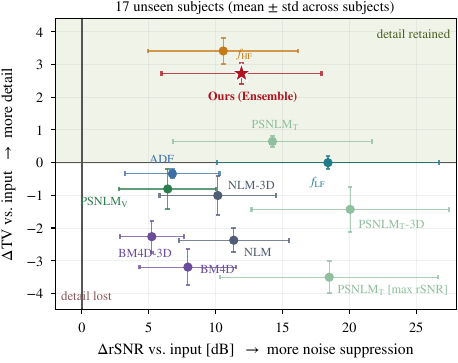}}
    \caption{Noise suppression against detail retention over the 17 test subjects. Each point is a method minus the noisy input of the same subject, and the bars are the standard deviation of that paired difference across subjects. All PSNLM points are the analytical filter alone, without the network.}
    \label{fig:quadrant}
\end{figure}

Table~\ref{tab:reffree} reports quality averaged over the 17 test subjects, and Fig.~\ref{fig:quadrant} the paired difference between each method and the input of that same subject. Fig.~\ref{fig:visual} illustrates one of these clips, with its own measurements printed underneath.

The two members and their fusion behave as designed. $f_{\mathrm{LF}}$ yields the highest rSNR of the three by prioritizing global smoothness, at the cost of TV and of an SSIM of 0.56 with respect to the input, which reflects structural deviation caused by over-smoothing. $f_{\mathrm{HF}}$ recovers sharper textures, with the largest TV gain among all methods considered, at an SSIM of 0.87. The ensemble balances the two effects, suppressing noise by 11.9~dB while increasing TV by 2.73 at an SSIM of 0.85, with visible improvements in fine articulatory structures such as the velum (Fig.~\ref{fig:visual}). Across the test subjects, the rSNR gain decreases and the TV gain increases monotonically with $w$, from $+$18.2~dB and $+$0.32 at $w=0.1$ to $+$11.2~dB and $+$3.08 at $w=0.9$.

With respect to the analytical denoisers ADF, NLM and BM4D, the latter two also in their 3-D forms, improve rSNR by between 5.2 and 11.4~dB, and reduce TV by 0.34 to 3.20 (Fig.~\ref{fig:quadrant}). $\mathrm{PSNLM}_{\mathrm{T}}$ achieves the largest rSNR gain among the analytical filters, 14.3~dB at the setting used to generate the pseudo-targets and 20.1~dB in its 3-D form. At that setting it is also the only analytical operating point in Fig.~\ref{fig:quadrant} that increases TV, by 0.65. Both results are obtained at an SSIM of 0.57, that is, with substantial deviation from the input. The Rician stabilized $\mathrm{PSNLM}_{\mathrm{V}}$ shows the opposite behavior: it maintains an SSIM of 0.80 but suppresses only 6.4~dB and still reduces TV. No analytical filter achieves the three properties simultaneously; only the ensemble and its $f_{\mathrm{HF}}$ member suppress noise, increase high-frequency content and remain close to the input. The additional gradient energy introduced by the ensemble remains correlated with the anatomy of the input with a 0.96 Pearson correlation, which indicates that it corresponds to recovered detail and not to added noise. The hyper-parameter search for the filters leads to the same conclusion: of the 5950 configurations scored on the tuning clips, 1870 match the structural fidelity of the ensemble, and none of these exceeds it in both rSNR and TV. The highest TV among them is 9.88, against 12.53 for the ensemble on the same clips, and no configuration in the entire search exceeds 10.27, for an input value of 9.59.

\subsection{Reference-Based Evaluation} \label{subsec:refbased}
\begin{figure}
    \centerline{\includegraphics[width=\linewidth]{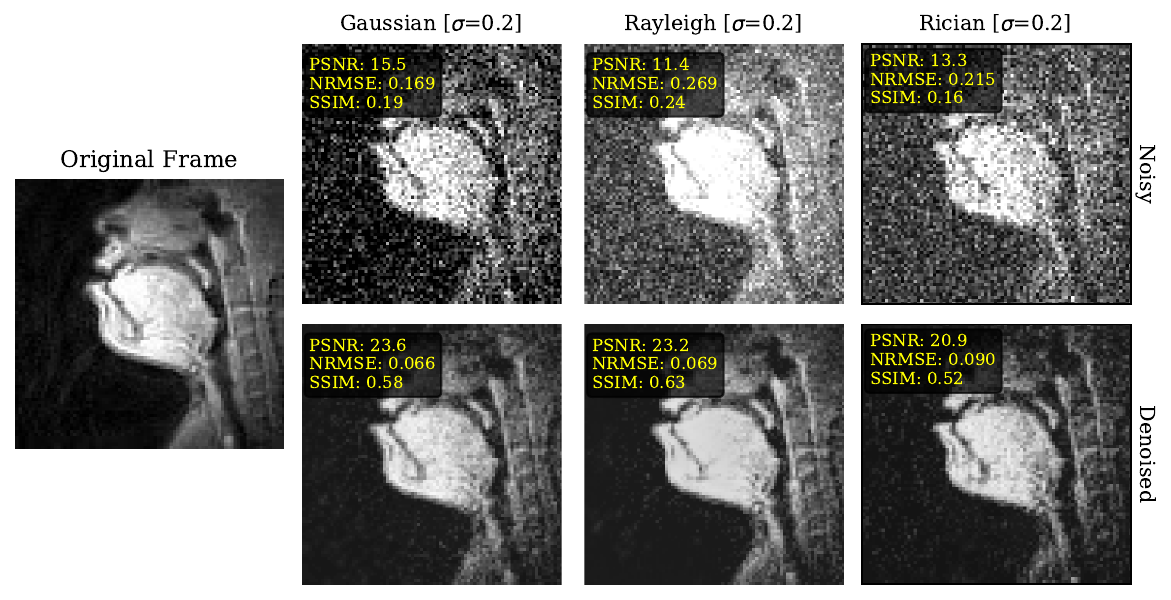}}
    \caption{The ensemble at $w=0.1$ on the original videos after Gaussian, Rayleigh and Rician noise is added at $\sigma=0.2$. Top: the corrupted input. Bottom: the output. The reference for the three metrics is the uncorrupted USC-75 frame on the left, which carries the scanner's own Rician noise.}
    \label{fig:sec}
\end{figure}

We additionally evaluate the ensemble on USC-75 frames corrupted with Gaussian, Rayleigh, and Rician noise at $\sigma=0.2$, using three reference-based metrics against the uncorrupted frame: the peak signal-to-noise ratio (PSNR),  the normalized root mean squared error (NRMSE) and SSIM, which here compares the output with that reference rather than with the noisy input. As shown in Fig.~\ref{fig:sec}, the method suppresses severe contamination under all three distributions, with $w=0.1$ to maintain the high SNR of $f_{\mathrm{LF}}$ while injecting sufficient high-frequency content to sharpen articulatory edges. PSNR increases from 15.5 to 23.6~dB under Gaussian noise, from 11.4 to 23.2~dB under Rayleigh noise and from 13.3 to 20.9~dB under Rician noise, and SSIM increases from 0.19, 0.24 and 0.16 to 0.58, 0.63 and 0.52 respectively, an approximately threefold improvement. This indicates that the geometry of the articulators is largely preserved under severe degradation.

\subsection{Pseudo-Supervision Ablation} \label{subsec:abl}

Both ensemble members were re-trained against three regression targets, the pseudo-target built with $F_{\mathrm{T}}$, the one built with $F_{\mathrm{V}}$ and the raw noisy clip, holding architecture, optimizer, schedule, data subset, seeds and number of updates fixed, so that the regression target is the only difference. These runs are shorter than those behind Table~\ref{tab:reffree} and their absolute values are correspondingly lower.

\begin{figure}[!h]
    \centerline{\includegraphics[width=\columnwidth]{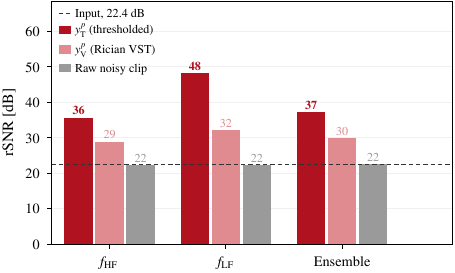}}
    \caption{Ablation of the pseudo-supervision with both ensemble members re-trained under a common setting except for the regression target. rSNR of $f_{\mathrm{HF}}$, $f_{\mathrm{LF}}$ and their fusion, trained against the thresholded pseudo-target $y^p_{\mathrm{T}}$, the Rician-VST pseudo-target $y^p_{\mathrm{V}}$, and the raw noisy clip, with the input's own rSNR as the dashed line.}
    \label{fig:ablation}
\end{figure}

The results are shown in Fig.~\ref{fig:ablation}. For $F_{\mathrm{T}}$ as pseudo-target the fused output reaches an rSNR of 37.1~dB, being reduced to 29.8~dB for $F_{\mathrm{V}}$, whereas training on the raw noisy clip gives 22.5~dB, indistinguishable from the 22.4~dB of its own input, at an SSIM of 0.98 with respect to that input. Without a filtered target the network does not denoise but reproduces its input. The sub-sampling of Sec.~\ref{subsec:ss} is therefore not sufficient on its own, since it decouples only the Charbonnier term of Eq.~\eqref{eq:loss} from the input noise. $\mathcal{L}_{G}$ is evaluated on all pixels, so with a noisy target it drives the output to reproduce the gradient field of the noise, and neither term addresses the non-zero-mean Rician floor of Eq.~\eqref{eq:rice}. The validation metrics of the original full-scale trainings, which paired the same configuration with and without the pseudo-target, are consistent with this: relative to its own input, the pseudo-target arm gains 11.8~dB and the noisy-target arm loses 0.7~dB, at essentially unchanged TV (7.87 and 7.70 respectively). The pseudo-target arm also exceeds the pseudo-target on which it was trained by 4.1~dB, showing that it does not simply reproduce the filter.


Between the two pseudo-targets, $F_{\mathrm{T}}$ yields an rSNR higher by 7.3~dB. Since $F_{\mathrm{T}}$ flattens the region that forms the denominator of rSNR, this metric alone is not conclusive. On the frame-varying component of the background, which a fixed pattern cannot reduce, the $F_{\mathrm{T}}$ arm lowers the standard deviation by a factor of 20.2, compared with 12.8 for $F_{\mathrm{V}}$ and 4.0 for the noisy target, while a control that only thresholds the input, without filtering, gives 0.9. The $F_{\mathrm{T}}$ arm also agrees better with the input structures inside the vocal tract giving a Pearson correlation of 0.95 against 0.87.

Table~\ref{tab:mechanism} shows which of the two operations composing $F_{\mathrm{T}}$, the hard threshold or the Anscombe root, accounts for its advantage over $F_{\mathrm{V}}$. This is a separate, smaller experiment than the one above: only the $f_{\mathrm{HF}}$ member is retrained, against a pseudo-target built from each half of Eq. \eqref{eq:thresh} in isolation, with the non-local means stage, the noise estimator, the data, the schedule and the seed held fixed as before. The threshold alone reaches 33.0~dB, most of the 35.6~dB obtained with the full transform, whereas the Anscombe root alone reaches 22.7~dB, the same level as the arm trained without a pseudo-target at all. The supervision therefore derives from the hard thresholding rather than from the variance stabilization alone for which the transform was initially introduced.

\begin{table}[!t]
\centering
\caption{Contribution of each component of the pseudo-target transform. Each row retrains $f_{\mathrm{HF}}$ against the threshold component, the Anscombe root component, or their combination $F_{\mathrm{T}}$, compared with the Rician-stabilized transform $F_{\mathrm{V}}$ and with no pseudo-target.}
\label{tab:mechanism}
\small
\setlength{\tabcolsep}{4pt}
\begin{tabular}{lccc}
\toprule
Pseudo-target & rSNR $\uparrow$ & $\Delta$ & SSIM \\
\midrule
Noisy input & 22.4 & n/a & 1.00 \\
\midrule
None (raw noisy clip) & 22.3 & $-0.1$ & 0.97 \\
Anscombe root only & 22.7 & $+0.3$ & 0.88 \\
$F_{\mathrm{V}}$, Rician VST & 28.9 & $+6.5$ & 0.92 \\
Threshold only & 33.0 & $+10.6$ & 0.88 \\
$F_{\mathrm{T}}$ (threshold $+$ root) & 35.6 & $+13.2$ & 0.86 \\
\bottomrule
\end{tabular}
\end{table}

\section{Conclusion}
We presented a self-supervised ensemble for blind vocal tract rtMRI denoising that requires no clean ground truth. By combining spatiotemporal SR training with a stabilized pseudo-supervision and a dual-model fusion strategy, the method effectively separates signal-dependent noise from dynamic articulatory structures across the USC-75 dataset. The ensemble mitigates the smoothness-sharpness trade-off inherent in single-model approaches. On the original data, a balanced fusion enhances high-frequency articulatory detail without distorting global anatomy, while under severe synthetic degradation, a stronger low-frequency contribution drives robust structural recovery. In both cases, the ensemble yields quantitatively and qualitatively superior results among the compared filters, demonstrating the viability of purely self-supervised methods for vocal tract rtMRI applications where clean reference data is physically unattainable. 

A comparison against different classical filters, searched exhaustively over their hyper-parameters, shows that suppressing the noise floor while increasing high-frequency articulatory content at comparable structural fidelity is not reachable by re-tuning the filters. The ablation shows that the pseudo-supervision, rather than the super-resolution constraint alone, drives the denoising, and that its effect comes from the hard threshold in $F_{\mathrm{T}}$ rather than from the variance-stabilizing root.

\section*{Acknowledgment}

We gratefully acknowledge the scientific support and computational resources provided by the Erlangen National High-Performance Computer Center (NHR@FAU) of the Friedrich-Alexander-Universität Erlangen-Nürnberg.

\balance
\bibliographystyle{IEEEtran}
\bibliography{refs}

\end{document}